\documentclass[12pt]{article}
\usepackage[english]{babel}
\usepackage[utf8x]{inputenc}
\usepackage{amsmath}
\usepackage{graphicx}
\usepackage{etoolbox}
\usepackage{changepage}
\usepackage{titlesec}
\usepackage[parfill]{parskip}
\usepackage[margin=1in]{geometry}
\usepackage{times}
\usepackage{float}
\usepackage[numbers]{natbib}
\usepackage{lipsum} 

\titleformat*{\section}{\normalsize\bfseries} 

\title{\large \textbf{Investigating Opportunities for Growth and Increased Diversity in Quantum Information Science and Engineering Education in the U.S. based on an Analysis of the Current Educational Landscape}}
\date{} 
\author{A.R. Pi\~na\textsuperscript{1}, Shams El-Adawy\textsuperscript{2,3}, Mike Verostek\textsuperscript{1},\\
H. J. Lewandowski\textsuperscript{2,3}, Benjamin M. Zwickl\textsuperscript{1}\\
\textit{School of Physics and Astronomy, Rochester Institute of Technology}\textsuperscript{1}\\
\textit{JILA, NIST and the University of Colorado}\textsuperscript{2}\\
\textit{Department of Physics, University of Colorado, Boulder}\textsuperscript{3}
}
\makeatletter 
\patchcmd{\@maketitle}{\begin{center}}{\begin{adjustwidth}{0.5in}{0.5in}\begin{center}}{}{}
\patchcmd{\@maketitle}{\end{center}}{\end{center}\end{adjustwidth}}{}{}
\makeatother

\begin{document}
\raggedright
\maketitle
\thispagestyle{empty}
\pagestyle{empty}

\section*{Abstract}

Quantum Information Science and Engineering (QISE) is rapidly gaining interest across a wide range of disciplines. 
As QISE continues to evolve, engineering will play an increasingly critical role  in advancing quantum technologies. 
While efforts to characterize introductory QISE courses are underway, a comprehensive understanding of QISE education across the United States remains lacking. 
Developing a broad understanding of the QISE education landscape is crucial for addressing the needs of the growing quantum industry and ensuring equitable access for a diverse range of participants.
This paper presents part of an ongoing effort to characterize the current landscape of QISE courses and degree programs in higher education in the US. 
To achieve this, we used publicly available information from university and college websites to capture information on over 8000 courses that address quantum in some way and nearly 90 QISE specific programs (e.g., degrees, minors, certificates). 
The majority of these programs are interdisciplinary and include engineering; 14 of them are housed exclusively in engineering departments. 
We find most programs are offered at research intensive institutions. 
Our results showcase an opportunity for program developers at non-research intensive institutions to justify the creation of QISE programs, which would also address calls from different stakeholders in QISE education for a more diverse QISE workforce.
We suggest strategies based on the findings of this study such as integrating QISE into existing engineering courses, investing in the development of QISE courses and programs at non-PhD-granting institutions, and making courses with QISE content accessible to students from a variety of majors.


\section*{Introduction}

In recent years, quantum technology has emerged as a federal priority driving investment in Quantum Information Science and Engineering (QISE) research and education. 
The National Quantum Initiative (NQI) Act was one of the first pieces of legislation in the US to establish the priority \cite{NQI}. 
Although it emphasized primarily the need for financial investment in research, the NQI act also calls for the establishment of a ``workforce pipeline''. 
The NQI act also established the Subcommittee on Quantum Information Science within the House Science, Space, and Technology Committee .
This subcommittee published the Quantum Information Science and Technology Workforce Development National Strategic Plan in 2022, which provides guidance on the development of such a ``pipeline'' and calls for higher education to support that development \cite{NSP}.
 A necessary step in continued development of courses and programs to support the QISE workforce pipeline is an assessment of the current state of QISE education \cite{CHIPS}.

Although quantum mechanics has traditionally been rooted in physics, QISE is inherently interdisciplinary, drawing significant contributions engineering, materials science, chemistry, optics, computer science, and mathematics. 
Engineering education plays a pivotal role in advancing QISE, prompting us to explore the following research questions:
\begin{enumerate}
    \item Which engineering subdisciplines offer the most quantum and QISE courses?
    \item What categories of engineering courses include quantum and QISE concepts?
    \item How many engineering courses mention specific QISE topics?
    \item How does course availability change with institutional characteristics?
\end{enumerate}
This paper builds on our previous work that explored a similar set of questions within the context of physics \cite{Pina_2025_landscape}.


\section*{Methods}

This project was conducted in three primary phases. 
The first phase involved institution selection, narrowing the 4,100 degree granting institutions in the US to those most relevant to the study. 
The second phase was focused on data collection, documenting details of quantum-related courses and programs. 
Finally, in the third phase, the data were categorized and organized to synthesize the information into the format presented here.

\textit{Institution selection}

We began by considering all institutions of higher education in the US, including two-year colleges and community colleges. Any institution that met at least one of the six following criteria was included in the sample for this study. 
 \begin{enumerate}
    \item Institutions classified as `Very High Research Activity' (N = 146) or `High Research Activity' (132) by the Carnegie Classification system.
    \item Top 10 STEM bachelors producing institutions in each State.
    \item Top 5 STEM associates producing institutions in each State.
    \item Institutions with ABET accreditation in computer science or engineering. (711)
    \item All Minority Serving Institutions (MSIs) (722)
\end{enumerate}

These criteria were chosen for several reasons,including the following.
\begin{itemize}
    \item Research-intensive institutions are the most likely to offer QISE courses, as noted by Meyer et al. \cite{Meyer2024}.
    \item Criteria three and four ensure representation from institutions in every State in the US, providing a geographically comprehensive sample.
    \item Criterion four specifically addresses a common shortcoming in studies of this nature by including two-year institutions, which are often overlooked.
    \item ABET accreditation ensures that institutions in the sample reflect the contributions of computer science and engineering to the development of QISE courses and programs.
    \item Minority Serving Institutions (MSIs) play a crucial role in advancing educational opportunities for historically marginalized populations, aligning with the goals of fostering diversity in QISE. Additionally, three legislative initiatives motivating this work—the National Quantum Initiative Act (NQI), the CHIPS Act, and the National Science and Technology Policy (NSP)—all emphasize the need for increased diversity in QISE \cite{NQI, CHIPS, NSP}.
 
\end{itemize}

All of the institutions included  in this study account for 98\% of engineering bachelors degrees awarded in the most recent year for which Integrated Postsecondary Education Data System \cite{IPEDS} data was available (2022) at the time of institution selection (June 2024). 
Therefore, we believe this study provides a robust and comprehensive overview of the integration of quantum topics within engineering curricula across the US.

\textit{Data collection}

Data were collected between June and September 2024 by a team of seven researchers resulting in approximately 600 total hours of effort. 
Each institution was investigated individually, beginning with a review of course and program catalogs. 
Course catalogs were searched for any courses with titles  or descriptions containing the the word ``quantum.'' 
For each course meeting this criterion, we recorded the title, description, course number, prerequisite course(s), department(s) associated with the course, and academic level(s). 
Similarly, programs with names containing the word ``quantum" were recorded, capturing details such as the home department, admission requirements, and program levels.
 Programs with names including the term ``quantum" were documented, capturing details such as the home department, admission requirements, and program levels.
To ensure consistency, all data were recorded using a standardized form using the Qualtrics survey platform. The collected data were then analyzed  and visualized using python code.

\textit{Department and course categorization}

Engineering is a broad discipline that typically  includes multiple departments at an institution, often with varied naming conventions for academic units. 
To make the data interpretable, it was necessary to group departments into broader categories. 
This grouping is particularly relevant for the categories of Electrical and Computer Engineering (ECE), Materials Science (and Engineering), and Other Engineering, which represent the majority of course data discussed in this paper.

Engineering departments were categorized into four major groups. A full listing of department names within each group can be found in an appendix of Ref.\cite{Pina2025_landscape}. Criteria for each group are:

\begin{itemize}
    \item \textbf{Electrical and Computer Engineering (ECE):} Any department name containing both ``computer'' and ``engineering'' or ``electrical'' and ``engineering'' was included in this category. Examples include ``Computer Systems Engineering,'' ``Computer Sciences Engineering,'' ``Electrical and Electronic Engineering,'' and ``Electrical Engineering and Computer Science.'' 
    
    \item \textbf{Computer Science:} Departments explicitly named ``Computer Science,'' but without ``Engineering'' in their title were categorized separately as Computer Science. ``Software Engineering'' was also grouped under Computer Science as a subset. 
    
    \item \textbf{Materials Science (and Engineering):} Departments with names including both ``materials'' and ``science'' or ``materials'' and ``engineering'' were grouped into this category. 
    
    \item \textbf{Other Engineering (Other Eng):} This category encompasses a wide range of departments, each of which individually contributes relatively few courses. Examples include ``Mechanical Engineering,'' ``Bioengineering,'' ``General Engineering,'' ``Chemical Engineering,'' ``Aerospace Engineering,'' and ``Nuclear Engineering.'' 
\end{itemize}

Courses required a similar categorization process. 
We began by conducting literal string searches within the titles of courses offered by engineering departments to identify thematic patterns. 
These searches facilitated the grouping of similar courses into preliminary categories.
For example, searching for the string "nano" yielded courses such as "Advanced Nanoelectronics," "Introduction to Nanoscience," "Fundamentals of Nanotechnology," and "Nanostructure Materials." 
While these courses vary in their specific applications or theoretical focus, they are unified by their emphasis on nanoscale systems.
Optics courses,  focused primarily on theoretical concepts, and photonics courses, emphasizing practical applications, were kept separate due to the clear distinctions in their titles and descriptions.

Using this method, approximately 90\% of the courses were grouped into 18 categories. 
The remaining uncategorized courses were reviewed manually, with their descriptions examined and categorized accordingly. 
All course descriptions were reviewed to ensure that each course was accurately assigned to its respective category.
Course descriptions were further analyzed for specific QISE topics identified by Meyer and colleagues \cite{Meyer2024}. 
We will discuss the data set in three primary subsets: all courses including `quantum' (the entire set of recorded courses), courses with QISE topics (courses with descriptions containing QISE keywords), and dedicated QISE courses. 
Within those subsets we will also look specifically at courses offered in engineering disciplines. 

\textit{Limitations}

These methods are not without limitations, and several challenges emerged during the process.
One significant issue stems from course catalogs themselves. 
At some institutions, course catalogs may lag behind actual changes implemented in classrooms, departments, or colleges. 
While a few widely used platforms exist for hosting online course catalogs, there is no standardization across institutions. 
For instance, some institutions provide a single, comprehensive catalog, while others maintain separate catalogs for undergraduate and graduate studies or even for individual disciplines.

Finding and navigating these catalogs can also pose challenges. 
They may be hosted on different parts of an institution’s website or have restricted access, available only to individuals affiliated with the institution. 
Additionally, course descriptions within catalogs are often vague, which can limit the specificity of claims about course content. 
While this vagueness provides flexibility for instructors to adapt or modify courses—particularly in cases where courses are taught by multiple instructors—it reduces the level of detail available for analysis.

Finally, during data collection and analysis, multiple rounds of binning and collapsing categories were necessary. While this approach improves interpretability, it inevitably results in the loss of some granularity and resolution in the data.


\section*{Results}
In our sample of 1,456 institutions, we identified 8,456 total STEM and non-STEM courses containing ``quantum" and 89 distinct degree programs in QISE. General findings from the full sample, including details about the QISE programs identified, can be found in \cite{Pina2025_landscape}.
Within engineering, we identified 14 distinct QISE programs, including undergraduate minors, master’s degree concentrations, full master’s degrees, concentrations/specializations, and Ph.D. programs. 

\textit{Engineering subdisciplines offering the most quantum and QISE coursework}

\begin{figure}[bt]
    \centering
    \includegraphics[width = \linewidth]{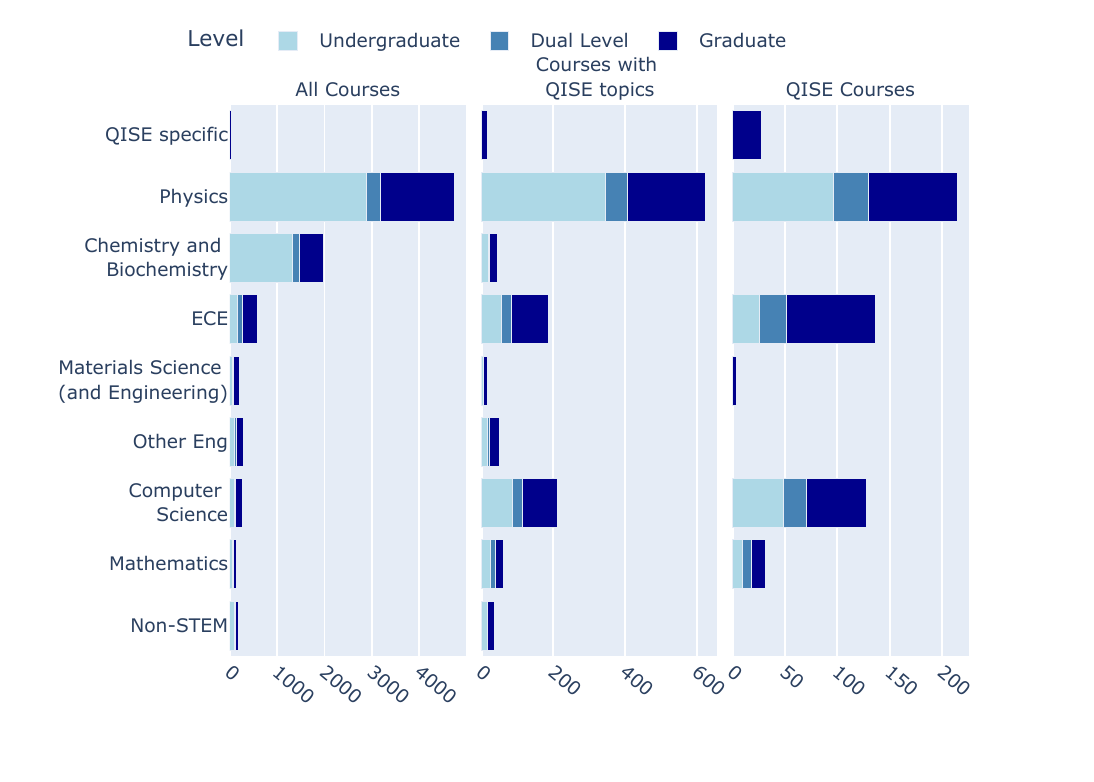}
    \caption{Number of courses at different levels offered in different disciplines from \cite{Pina2025_landscape} that contain quantum or QISE topics. Physics, chemistry, and ECE offer the most courses containing quantum concepts (left). QISE topics are most common in Physics, ECE, and CS (middle), which also offer the majority of QISE specific courses (right). }
    \label{fig:courses_per_discipline}
\end{figure}

The courses we recorded spanned a range of disciplines and were offered at three primary academic levels (see Fig.~\ref{fig:courses_per_discipline}).
Approximately 60\% of the recorded courses were listed in physics departments, while about 24\% were listed in chemistry and biochemistry. Engineering departments contributed 1,058 courses, accounting for approximately 12\% of the total, making it the third-largest discipline represented in this dataset.

The engineering courses identified were distributed across 226 institutions in the sample, representing approximately 34\% of all institutions with ABET accreditation in engineering or computer science. 
Even excluding other engineering fields, Electrical and Computer Engineering (ECE) offers the third-largest collection of courses containing ``quantum."

Despite the inclusion of other engineering fields, such as mechanical engineering (one of the largest subdisciplines), chemical, and biomedical engineering, these areas collectively account for only 284 courses.

ECE also stands out as a leading discipline in offering both courses that include QISE topics and QISE-specific courses, alongside computer science and physics. 
One notable distinction, as shown in Fig. \ref{fig:courses_per_discipline}, is the level at which these courses are offered.
In physics, the majority of ``quantum" courses are at the undergraduate level, whereas most engineering courses are offered at the graduate level.

The following sections will provide a closer examination of the categories of engineering courses that include ``quantum," the prevalence of QISE topics within these courses, and the availability of courses based on specific institutional characteristics.

\textit{Categories of engineering courses covering quantum and QISE topics}

\begin{figure}[htb]
    \centering
    \includegraphics[width = .9\linewidth]{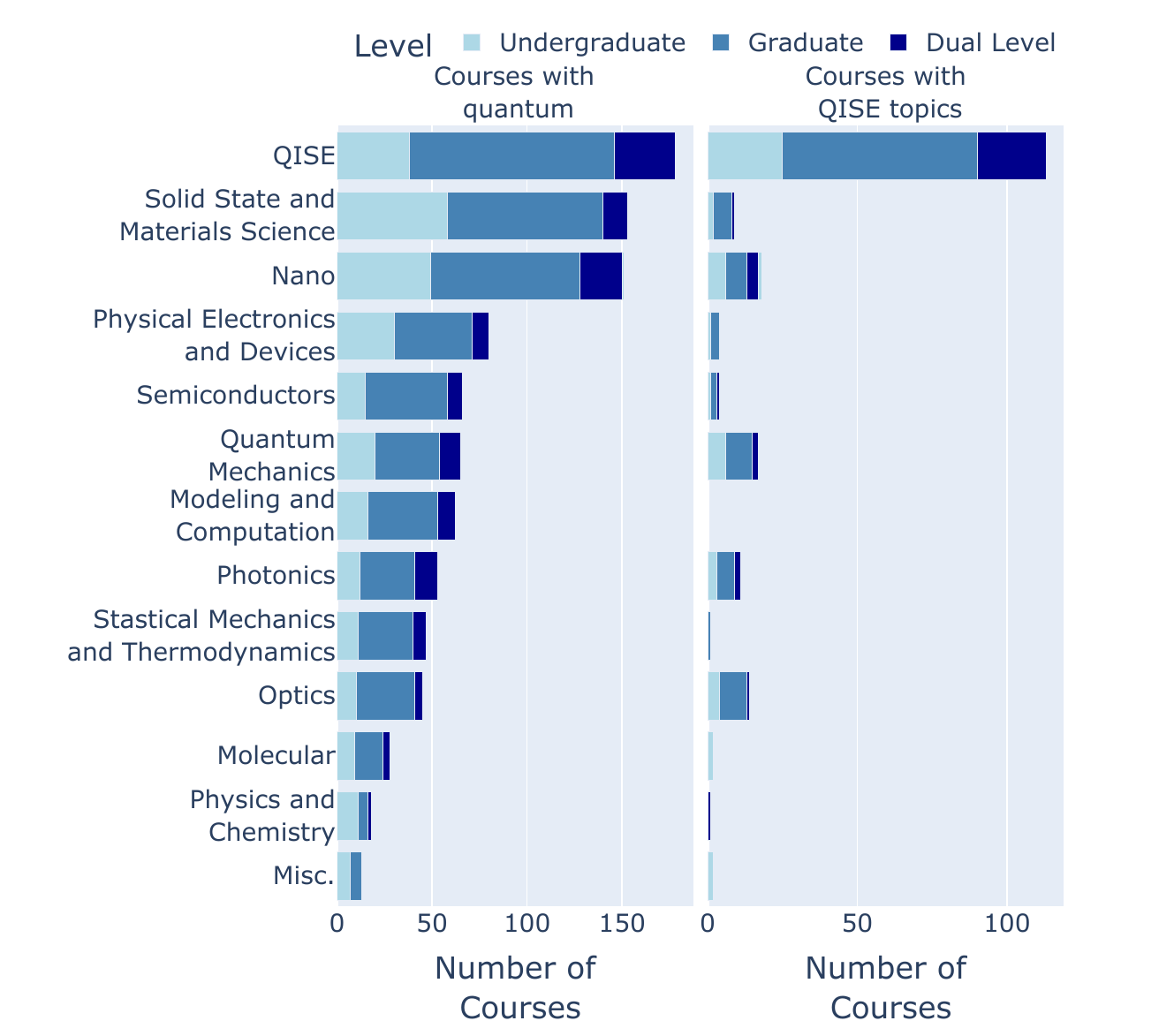}
    \caption{A summary of engineering courses from different categories with titles or descriptions that include ``quantum" (left) and QISE topics (right), colored by the level at which the courses are offered.}
    \label{fig:eng_courses_with_quantum}
\end{figure}

The course categorization described in the methods section allowed us to analyze related groups of   engineering courses that include ``quantum" in their title or description. 
Figure~\ref{fig:eng_courses_with_quantum} (left) shows a breakdown by engineering course category of these 1,058 `quantum' courses. Figure~\ref{fig:eng_courses_with_quantum} (right) shows 256 courses with descriptions that include one or more topics such as quantum computing and information, quantum devices, quantum sensing, quantum engineering, quantum electronics, and quantum cryptography.

QISE courses represent the largest single category among both engineering courses that mention ``quantum" and those that explicitly include QISE topics.
Figure~\ref{fig:eng_courses_with_quantum} (right) shows that QISE topics are not often included in any one category of courses outside of QISE. 
All together however, 116 of the 256 courses covering QISE topics are not QISE-specific courses. 
Certain categories of courses could serve as natural venues for integrating QISE topics into existing curricula; the fact that 45\% of courses covering QISE topics aren't categorized as QISE courses shows that this is already occurring to some extent.

Comparing the left and right plots in Fig. \ref{fig:eng_courses_with_quantum}, two categories, quantum mechanics and nano-focused courses, stand out for the extent to which non-QISE courses incorporate QISE topics. These courses demonstrate how QISE concepts are being integrated into broader engineering disciplines, highlighting areas where QISE topics are already gaining traction.
The nano- courses are particularly interesting because within those are the majority of courses that explicitly mention quantum sensing.
Although quantum sensing represents a significant portion of modern quantum technologies, it is not often part of QISE instruction.

\textit{Number of engineering courses that mention specific QISE topics}

\begin{figure}[ht]
    \centering
    \includegraphics[width = .9\linewidth]{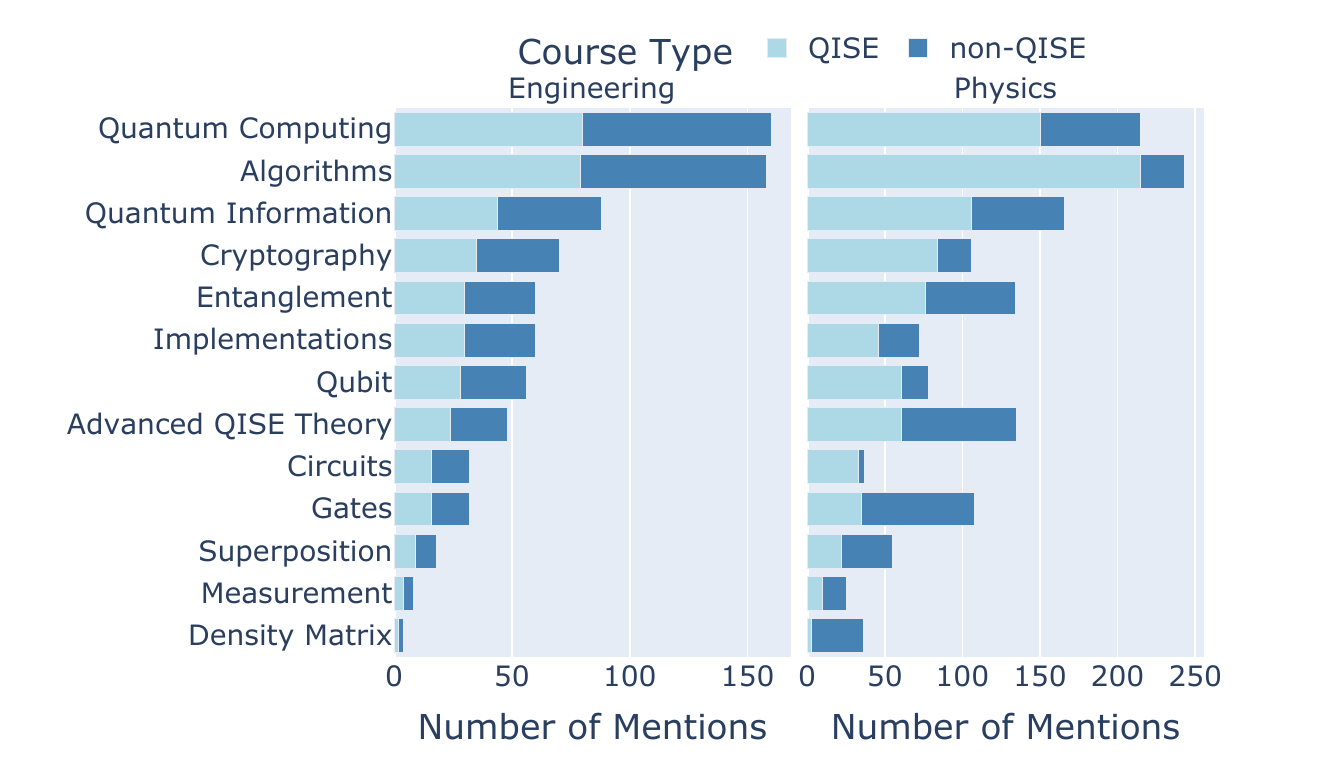}
    \caption{Counts of engineering courses teaching QISE topics. QISE refers to courses specifically focused on to QISE. Non-QISE includes all other recorded engineering course.}
    \label{fig:qise_topics}
\end{figure}

The same string searches that allow us to identify courses covering QISE topics allowed us to examine which QISE topics are most common in those courses. 
Counts of course descriptions mentioning different QISE topics are shown in Fig. \ref{fig:qise_topics}.
Note that the bars do not add to the total number of courses identified covering QISE topics (256) due to the fact that some courses mention multiple topics in their descriptions. 
Quantum algorithms and computing are the two most commonly covered QISE topics in engineering courses. 
In contrast, density matrix representations and quantum measurement are not commonly covered in engineering courses.

\textit{Course availability based on institutional characteristics}

A key perspective for assessing course availability is to examine the number of courses offered at different types of institutions (Two-Year, Four-Year, Masters-granting, PhD-granting). Table \ref{tab:avg_courses_per_carn} provides a summary of the number of engineering courses recorded for institutions with different Carnegie classifications.
These  courses per institution are calculated for the subset of institutions with ABET-accredited engineering or computer programs, as institutions without such programs are unlikely to offer engineering courses.
\begin{table}[htb]
    \centering
    \begin{tabular}{c|c|c|c}
        \textbf{Institution Type} & \textbf{Number of Institutions} & \textbf{Quantum} & \textbf{QISE} \\
        \hline
        \textbf{Two-Year} & 72 & 0.17 & 0 \\
        \textbf{Four-Year} & 90 &  0.12 & 0.01 \\
        \textbf{Masters} & 176 & 0.31  & 0.07 \\
        \textbf{PhD} & 293 & 2.74  & 0.56 \\
    \end{tabular}
    \caption{Availability of quantum and QISE-related coursework across institution types.  The ``Number of Institutions" refers to the institutions in the sample that have an ABET-accredited engineering or computer science program. The latter two columns denote the number of courses per institution. ``Quantum'' denotes any course with a title or description that includes the term ``quantum," while ``QISE'' refers to courses with descriptions covering one or more QISE topics as defined in Fig. \ref{fig:eng_courses_with_quantum}.}
    \label{tab:avg_courses_per_carn}
\end{table}

Students at two-year and four-year institutions have limited access to engineering courses mentioning ``quantum" or covering QISE topics. 
Because these institutions typically have 0 or 1 courses quantum-related courses, roughly 17\% of two-year colleges and 12\% of four-year colleges contain a quantum-related engineering course, and QISE-related courses are nearly completely absent within engineering. 
On average, PhD-granting institutions offer almost three quantum-related engineering courses, and about half of PhD-granting institutions have an engineering course covering QISE topics.

There are 155 MSIs with at least one ABET accredited program in engineering or computer science, 67 of which have courses that mention ``quantum.''
Similar to the set of all institutions presented in Table~\ref{tab:avg_courses_per_carn}, Table~\ref{tab:avg_courses_per_carn_msi} shows that at two-year and four-year institutions are unlikely to have courses including ``quantum'', and none of them have a course covering QISE topics.
Masters and PhD-granting MSIs are more likely to have a course including  ``quantum'' or QISE topics
so these institutions have more courses covering these topics. 
Of the 67 MSIs  with courses, 45 are are classified as either ``high research activity'' or ``very high research activity''. 
Comparing Table \ref{tab:avg_courses_per_carn} to Table~\ref{tab:avg_courses_per_carn_msi} also shows that there are both more quantum and QISE courses per institution at the masters and PhD-granting MSIs than the there are in the set of all institutions. 
The differences based on institution type as well as comparisons between MSIs and non-MSIs suggest that a finer grained analysis looking at different subsets of MSIs may be necessary
(e.g., a rural HBCU is very different from an urban very high research activity HSI). 

\begin{table}[htb]
    \centering
    \begin{tabular}{c|c|c|c}
        \textbf{Institution Type} & \textbf{Number of Institutions} & \textbf{Quantum} & \textbf{QISE} \\
        \hline
        \textbf{Two-Year} & 20 & 0 & 0 \\
        \textbf{Four-Year} & 13 &  0.08 & 0 \\
        \textbf{Masters} & 40 & 0.73  & 0.25 \\
        \textbf{PhD} & 72 & 3.69  & 0.72 \\
    \end{tabular}
    \caption{Availability of quantum and QISE-related coursework across MSIs of different Carnegie classifications.  The ``Number of Institutions" refers to the MSIs in the sample that have an ABET-accredited engineering or computer science program. The latter two columns denote the number of courses per institution. ``Quantum'' denotes any course with a title or description that includes the term ``quantum," while ``QISE'' refers to courses with descriptions covering one or more QISE topics as defined in Fig. \ref{fig:eng_courses_with_quantum}.}
    \label{tab:avg_courses_per_carn_msi}
\end{table}


\section*{Discussion}

Although quantum is often considered  exclusively a subset of physics, our data collection and analysis show that quantum mechanics content is present in engineering courses. 
This is especially true for ECE, which, among the subdisciplines of engineering, is much more prominent in courses including ``quantum'' and QISE. 
QISE utilizes technology and techniques that are already ubiquitous in ECE, making ECE an ideal place to incorporate QISE topics into existing instruction. 
Examples of ECE courses that integrate QISE include computer architecture courses covering quantum computing hardware, cybersecurity courses addressing quantum cryptography, and nanotechnology courses exploring quantum sensing platforms. 

Solid state, materials science, and semiconductor courses also have some coverage of quantum, but little coverage of QISE.
This gap presents a unique opportunity, as one of the leading platforms for quantum computing is superconducting qubits, which relies heavily on expertise in materials science and related disciplines. 
Additionally, quantum simulations of materials are one of the most promising application areas for quantum computing hardware.
Integrating QISE into courses in materials science and engineering could bridge this knowledge gap and prepare students to contribute to advancements at the intersection of quantum technologies and materials engineering.

Another subdiscipline, mechanical engineering, which awards the largest number of bachelor's degrees of all engineering subdisciplines \cite{ASEE_ENG_by_numbers}, is noteworthy as it accounts for only 108 ``quantum'' courses across the entire sample and only 1 QISE course.
Although quantum and QISE may have less overlap with mechanical engineering, there may be interest related to quantum sensing of motion for navigation and quantum imaging that are relevant for robotics and mechatronics. 
Given the large number of students in mechanical engineering programs, the subdiscipline cannot be overlooked as a possible venue to introduce engineering students to quantum and QISE. 

Examining the academic level at which these courses are taught, we find that both quantum and QISE content in the context of engineering is more prevalent at the graduate level.
Although there are 38 undergraduate engineering QISE courses in our sample with titles such as ``Quantum Computing,'' ``Introduction to Quantum Technologies,'' and ``Quantum Cryptography,'' they represent a fraction (27\%) of QISE courses in engineering. 
One possible way to increase students' access to these courses could be making them interdisciplinary, so engineering students from different subdisciplines could enroll. 

Another way to examine the integration of QISE into engineering curricula is through comparing engineering to the discipline offering the most quantum courses, physics. Within physics, quantum ideas are present in two courses common to nearly all physics degree programs: quantum mechanics and modern physics. These courses are typically required for all physics majors. Within engineering, there are substantially fewer courses with quantum content, suggesting these courses are not required for most engineering majors or quantum ideas may be optional parts of a course. The most common categories of courses including ``quantum'' include QISE, Solid State and Materials Science, and Nano. Although teaching general ideas about quantum is less prevalent in engineering, many dedicated QISE courses are being offered in engineering, particularly ECE. In fact, QISE courses were the largest single category of courses teaching quantum ideas within engineering and the total number of QISE courses in engineering is not far behind physics (see Fig.~\ref{fig:courses_per_discipline}).  

We now consider the types of institutions at which these courses are offered. 
Engineering courses covering both quantum generally, as well as QISE specifically, are concentrated at PhD-granting institutions. 
Most non-PhD-granting institutions offering engineering degrees do not have a single engineering course that mentions ``quantum" in its course description, whereas PhD-granting institutions typically have two or three courses, which  highlights disparities in the availability of quantum-related educational opportunities based on institution type.
This disparity is consistent with previous studies on the availability of QISE courses \cite{Cervantes2021, Meyer2024,Pina2025_landscape}.
In the US, there are more non-PhD-granting than PhD-granting institutions.
Non-PhD-granting institutions collectively reach both more students and a more diverse population of students, making them a good place for targeted investment and development of QISE programs and coursework. 

Finally, we consider how QISE courses might integrate into engineering degree programs.
Engineering degrees are among the most structured and least flexible academic programs, with limited room for elective coursework. 
This rigidity often stems from the broad scope of knowledge and practical skills required for engineering students to succeed.
As a result, adding standalone QISE courses to all existing engineering programs may not always be feasible. 
However, the strong connections between quantum theory, quantum technologies, and the established engineering fields present opportunities for integration. 
Embedding QISE concepts into relevant courses (e.g., physical electronics, materials science, or semiconductors)  or modules within these disciplines could enhance students’ understanding while minimizing the need for additional coursework. 
This approach has the potential to equip students with skills and insights that open new opportunities in both academia and industry. 
Furthermore, it could contribute to building a more quantum-aware engineering and technology workforce, aligning with the growing demand for expertise in quantum information science and engineering. 

\section*{Conclusion}
In this paper, we overview the landscape of QISE courses and degree programs in engineering within higher education in the US. 
Our analysis highlighted that quantum and QISE content in engineering is more likely to be found in graduate-level courses and at PhD granting institutions.

As the demand for a quantum workforce increases, our results showcase a few opportunities where engineering programs and courses could consider incorporating QISE:
\begin{itemize}
    \item Embedding QISE concepts or modules into relevant existing courses within engineering disciplines
    \item Investing in the development of QISE programs and coursework at non-PhD-granting institutions
    \item Ensuring access to QISE content across a wide variety of engineering subdisciplines by offering interdisciplinary courses  
    \item Considering opportunities to integrate QISE into the mechanical engineering, which is the largest engineering subdiscipline by degrees granted
\end{itemize}
Furthermore, we hope that this paper can motivate and guide the creation of new QISE curricular materials, courses, and programs in engineering.


\vspace{4\baselineskip}\vspace{-\parskip} 
\footnotesize 
\bibliographystyle{ieeetr} 


\end{document}